\documentstyle[nato]{crckapb} 
\input epsf 
\newcommand{\beq}{\begin{equation}}
\newcommand{\eeq}{\end{equation}}
\newcommand{\bea}{\begin{eqnarray}}
\newcommand{\eea}{\end{eqnarray}}

\newcommand{\tr}{{\rm tr}}
\newcommand{\V}{{\cal V}}
\newcommand{\vev}[1]{\Big\langle #1 \Big\rangle}

\begin{opening}
\title{VORTICES AND CONFINEMENT}

\author{T.G. KOV\'ACS}
\institute{Instituut-Lorentz for Theoretical Physics, \\
P.O.Box 9506, 2300 RA, Leiden, The Netherlands}
\author{E.T. TOMBOULIS}
\institute{Department of Physics, UCLA, \\
Los Angeles, CA 90095-1547, USA}

\end{opening}

\begin{document}

\begin{abstract}   
We review recent developments in the vortex picture of confinement.  
We discuss numerical simulations demonstrating that the entire asymptotic 
string tension is due to vortex-induced fluctuations of the Wilson loop. 
Analytical and numerical results concerning the presence of vortices as the necessary and sufficient condition for confinement at arbitrarily weak 
coupling in SU(N) gauge theories are also discussed.\footnote{Presented 
by E.T. Tomboulis at the workshop ``Lattice fermions and structure of the 
vacuum'', 5-9 October 1999, Dubna, Russia.}   
\end{abstract}

\section{Introduction} 

The proposal that extended vortex configurations are responsible 
for maintaining confinement at arbitrarily weak coupling in $SU(N)$ gauge 
theories has a long history \cite{tH} - \cite{Yo}. Several important 
results were established by the early eighties. Over the last two 
years, the vortex picture of confinement has undergone very 
substantial development by a series of numerical investigations as 
well as new analytical results \cite{lat}-\cite{forc}.

It was originally conjectured that thick vortices 
occur with nonvanishing measure contribution in the path integral at 
arbitrarily large beta, and that this provides a sufficient mechanism 
for confinement. In light of recent developments, it now appears 
that not only is this contribution sufficient but 
also necessary: it is responsible for the full string tension 
of large Wilson loops in $SU(N)$ gauge theories. After briefly 
recalling basic features of the vortex picture, we  
discuss some of the recent numerical and analytical results 
underlying these developments. Conclusions are presented in section 
\ref{concls}.

\section{Physical Picture} 

A vortex configuration of the gauge field may be characterized  
by a multivalued singular $SU(N)$ gauge transformation function 
$V(x)$. The multivaluedness ambiguity lies in the center $Z(N)$, 
so the transformation is single-valued in $SU(N)/Z(N)$. If one 
attempts to extend such a gauge transformation $V(x)$ throughout 
spacetime, it becomes singular on  a closed surface $\V$ of 
codimension 2 (i.e. a closed  loop in $d=3$, a closed 
2-dimensional sheet in $d=4$) forming the topological obstruction 
to a single-valued choice of $V(x)$ throughout spacetime. Generic 
vortex configurations of the gauge potentials then consist of a  
pure-gauge long-range tail given by $V(x)$, and a core enclosing 
the region where $\V$ would be if one were to try to 
smoothly extend $V$ everywhere. Equivalently, the configuration 
cannot be smoothly deformed to pure gauge everywhere without 
encountering a topological obstruction $\V$. Note that 
it is only the existence of this obstruction, and not its precise 
location, that is relevant; the location may always be 
moved around by a regular gauge transformation. The 
asymptotic pure-gauge part provides then a topological 
characterization of the configurations irrespective 
of the detailed structure of the core.

Assume now that two gauge field configurations  $A_\mu(x)$ 
and $A^\prime_\mu(x)=V A V^{-1} + V\partial_\mu V^{-1}$ differ
by such a singular gauge transformation $V(x)$, and denote 
the path ordered exponentials of $A_\mu$ and $A^\prime_\mu$   
around a loop $C$ by $U[C]$ and $U^\prime[C]$, respectively.  
Then $\tr\,U^\prime[C]=z\,\tr\,U[C]$, where $z\neq1$ is a 
nontrivial element of the center,
whenever $V$ has obstruction $\V$ linking with the loop 
$C$; otherwise, $z=1$. Conversely, changes in the value 
of $\tr U[C]$ by elements of the center can be undone by 
singular gauge transformations on the gauge field 
configuration linking with the loop $C$. This means that 
vortex configurations are topologically characterized by 
elements of $\pi_1(SU(N)/Z(N))=Z(N)$. This topological $Z(N)$ 
flux is of course conserved only mod $N$, 
and, hence, the number of vortices in a given gauge field 
configuration in a given spacetime region can be 
defined only mod $N$.    

Vortex configurations, if sufficiently spread out, can be present  
at any nonzero coupling. This is because,  
by spreading the flux over a sufficiently  `thick' core,   
one can incur suffiently small cost in local action so that the 
configuration is not energetically suppressed at large beta; 
while at the same time there is very 
substantial disordering over long distances. In this way UV 
asymptotic freedom can coexist with IR confinement. 
The crucial question then is whether this class of configurations 
contributes with enough weight in the path integral measure 
to provide a sufficient mechanism for confinement at large beta. 
One may suspedt that this is so since it is easily seen that, 
given one vertex configuration, 
an enormous number of others may be produced by fluctuations 
that do not alter the vorticity content.

The vortex picture of confinement may then be summarized 
as follows. Confinement at arbitrarily weak coupling in 
$SU(N)$ gauge theories is the result of {\it nonzero vorticity 
in the vacuum over sufficiently large scales}. In other words, 
in any suffiently large spacetime region, the expectation 
for the presence of a spread out vortex is nonzero at all 
large $\beta<\infty$. This is strikingly  
demonstrated by a recent lattice computation     
which shows that the relative probability for vortex excitation 
in fact approaches unity (section \ref{Vex}). Configurations carrying 
sufficiently thick vortices contribute 
with essentially the same weight as ones without vortices. In 
this sense one has a `condensate' of vortex configurations.  
 
On the lattice a discontinuous (singular) gauge  
transformation introduces a {\it thin vortex}. The topological obstruction 
$\V$ is regulated to a coclosed set of plaquettes (in $d=4$ this is 
a closed 2-dimensional surface of dual plaquettes on the dual 
lattice). This represents the core of the thin vortex, each 
plaquette in $\V$ carrying flux $z\in Z(N)$.   

{\it Thick vortex} configurations can be constructed by
perturbing the bond variables $U_b$ in the boundary of each 
plaquette $p$ in $\V$ so as to cancel the flux $z$ on $p$, and 
distribute it over the neighboring plaquettes. Continuing this 
process by perturbing bonds in the neighboring $p$'s 
one may distribute the flux over a thickened core in the 
two directions transverse to $\V$. Beyond the thickness of 
the core, the vortex 
contribution reduces to the original multivalued pure gauge. 
If the original thin vortex is long enough, it may be
made thick enough, so that each plaquette receives a 
correspondingly tiny portion of the original flux $z$ that 
used to be on each $p$ in $\V$. Long thick vortices may 
therefore be introduced in $\{U_b\}$ 
configurations having $\tr U_p\sim \tr 1$ for all $p$. 
(Here $U_p$ denotes the product of the $U_b$'s around the 
plaquette boundary.) Thus they may survive at weak coupling 
where the plaquette action becomes highly peaked around 
$\tr U_p\sim \tr 1$. Long vortices may link with a large Wilson loop 
anywhere over the area bounded by the loop, thus potentially 
disordering the loop and leading to confining behavior. 
Thin vortices, on the other hand, necessarily incur a   
cost proportional to the size of $\V$, and only short ones 
can be expected to survive at weak coupling. These can link 
then only along the perimeter of a large loop generating only 
perimeter effects.

\newpage

\section{Isolation of Vortices and their Contribution - 
Simulation Results} 

Considerable activity has been devoted over the last two years 
to the isolation of vortices, and the computation of their 
contribution to the heavy quark potential and other physical 
quantities on the lattice. Several approaches have been pursued:
\begin{itemize} 
\item Computation of the expectation of the Wilson loop fluctuation  
solely by elements of the center on smoothed configurations which 
remove short distance fluctuations. 
\item Center projection in maximal center gauge (MCG) 
and isolation of P-vortices.  
\item Direct computation of excitation probability of a vortex in 
the vacuum (magnetic-flux free energy).
\end{itemize} 

Of these, the third is the most direct and physically transparent, 
and closely connected to the formulation of rigorous analytical 
results on the necessity and sufficiency of vortices for confinement. 
It also is computationally the most expensive. The result of a recent 
computation is presented in section \ref{Vex} below. In this section 
we discuss the first two. 

\subsection{Quark potential from center fluctuations on smoothed 
configurations} 
\label{smooth} 

We saw above that the fluctuation in the value of $\tr U[C]$ by 
elements of $Z(N)$, parametrizing the different $\pi_1(SU(N)/Z(N))$ 
homotopy sectors, 
expresses the changes in the number (mod $N$) 
of vortices linked with the loop over the set of configurations 
for which it is evaluated. One is not, however, interested in 
fluctuations produced by small thin vortices, which will still be  
present even at large beta but are irrelevant to the long distance 
physics. One would like to isolate only fluctuations 
by elements of the center due to extended configurations  
that can reflect long distance dynamics. To ensure 
this one performs local smoothing on the configurations 
which is constructed so that it removes short distance 
fluctuations  but preserves long distance physics. 
There is another, in fact essential reason for employing smoothing:  
it ensures good topological representation of extended fluctuations 
on the lattice.   
(According to rigorous theorems, only for lattice 
configurations with sufficiently small variations of the plaquette 
function $U_p$ from its maximum is it possible to 
unambiguously define a continuum 
interpolation assignable to a topological sector.)   

\begin{figure}[htb]
\begin{center}
\leavevmode
\epsfxsize=85mm
\epsfysize=85mm
\epsfbox{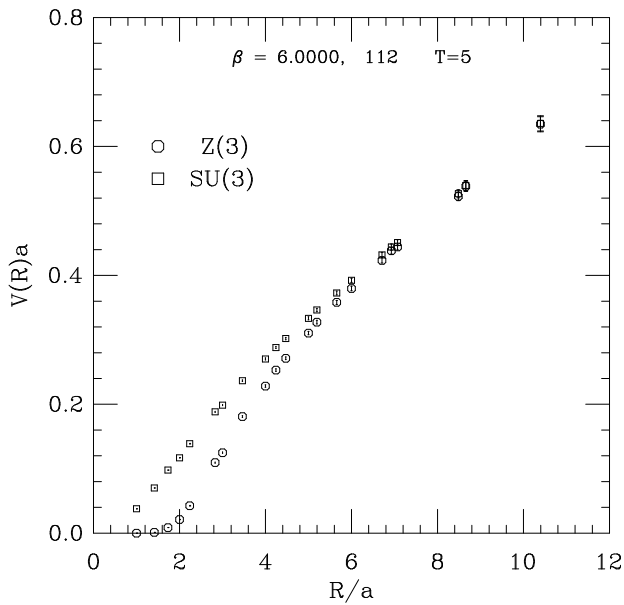}
\end{center}
\vspace{-0.7cm}
\caption{The heavy quark potential at $\beta=6.0$ on a set of
112 $12^3*16$ lattices extracted at time slice T=5 from 6 times
smoothed lattices.}
   \label{fig:potsu3_b6.0_b6_t5}
\end{figure}

Separate out the $Z(N)$ part of the Wilson loop observable 
by writing $\arg (\tr U[C]) = \varphi[C]+ {2\pi \over N}n[C]$, 
where $-\pi/N < \varphi[C]\leq \pi/N$, and $n[C]=0,1,\ldots,N-1$. 
Thus, with $\eta[C]=\exp(i{2\pi \over N}n[C]) \in Z(N)$, 
\bea 
W[C] = \vev{\tr U[C]} 
          & = & \vev{|\tr U[C]|\,e^{i\varphi[C]}\,\eta[C]} \nonumber\\
       &=& \vev{ |\tr U[C]|\,\cos(\varphi[C])\,
              \cos({2\pi \over N}n[C])}\;, \label{WL}
\eea 
using the fact that the expectation is real by reflection 
positivity, and that it is invariant under $n[C] \to (N-n[C])$. 
Next define 
\beq 
W_{Z(N)}[C] = \vev{\cos({2\pi \over N}n[C])}\;. \label{ZnWL}
\eeq 
One then compares the string tension extracted from the 
full Wilson loop $W[C]$, eq. (\ref{WL}), to the string tension 
extracted from $W_{Z(N)}[C]$, eq. (\ref{ZnWL}), on sets of 
progressivelly smoothed configurations \cite{KT1}-\cite{KT2}.  
Results have been obtained for $N=2$ and $N=3$ using the smoothing 
procedure in \cite{DeGet}. Typical results for the heavy 
quark potential for $N=3$ \cite{KT2} from six times smoothed lattices 
are shown in figure \ref{fig:potsu3_b6.0_b6_t5}.  
There is striking coincidence of the potentials extracted from 
(\ref{WL}) and (\ref{ZnWL}) at large distances indicating that 
the full asymptotic string tension is carried by the 
vortex-induced center fluctuations. This  coincidence is 
found to be very robust and stable under different number of 
smoothings. This is a very stringent test, as configurations 
subjected to different degrees of smoothings are vastly  
different, and indicates that this is an actual long-distance 
physics effect.

\subsection{Center projection, MCG and P-vortices} 
Most of the numerical simulations follow this approach. As  
almost all results in MCG are 
for $N=2$, we discuss only this case. The method \cite{Get} 
consists of the following steps. 

1. Fix the gauge by maximizing the quantity: 
$\quad \sum_b |\,\tr U_b\,|^2 \;$. 
This is the MCG. 

2. Make the center projecction: $\quad U_b \to Z_b\;$  
by replacing each $SU(2)$-valued bond variable by the 
closest center element $Z_b$. 

3. The excitations of the resulting $Z(2)$ bond configurations 
are coclosed sets of plaquettes each carrying $-1$ flux, i.e. 
$Z(2)$ vortices. These are the projection vortices (P-vortices).

The string tension extracted from loops of the center projected  
Z(2) variables is then found \cite{Get}, \cite{tub} to reproduce the full 
asymptotic string tension of the SU(2) LGT.  

The rational for the method is as follows.  Consider two configurations 
of the bond variables $U$ and $U'$ that differ by a discontnuous 
(singular) gauge transformation introducing a vortex. The corresponding  configurations in the adjoint representation $U_A$ and $U_A^\prime$ are then 
gauge equivalent by a regular gauge transormation. Now go to MCG.  
Then $U_A$ and $U_A^\prime$ go to the same MCG-fixed adjoint 
configuration $\bar{U}_A$; whereas $U$ and $U^\prime$ are transformed 
to $\bar{U}$ and $\bar{U}^\prime$ corresponding to the same adjoint 
$\bar{U}_A$. Hence $\bar{U}$ and $\bar{U}^\prime$ can only differ by 
a discontinuous plus possibly regular $Z(2)$ gauge transformations. 
Upon center projection, the projected $Z$ and $Z^\prime$ configurations 
will also differ by the same discontinuous $Z(2)$ transformation 
(plus possibly regular transformations), i.e 
one P-vortex (mod $N$) reflecting the vortex introduced 
by the singular transformation by which $U$ and $U^\prime$ differ. 

Note that the projected configurations $Z$ and $Z^\prime$ give 
for any Wilson loop linking with the P-vortex values differing by a 
sign (nontrivial element of $Z(2)$). But, as we saw in the previous section,  
the value of the Wilson loop in the original $U$ and $U^\prime$ 
will indeed differ by a sign if they differ by a discontinuous gauge 
transformation. In this way contact is made with the method in 
\ref{smooth} above.  

The above, however, relies on certain assumptions. It assumes that: 
a) $U_A$ and $U_A^\prime$ are gauge equivalent everywhere; 
b) the gauge fixing of the adjoint links by the MCG is complete 
everywhere leaving only a residual $Z(2)$ symmetry. It turns out, 
however, that, for the purpose of associating a P-vortex with a thick 
vortex, these assumptions cannot hold in general. First it is 
evident that a) applies strictly only if the singular gauge transformation 
corresponds to a {\it thin} vortex; for a {\it thick} vortex it cannot apply 
in the thick core, hence there is no unique $\bar{U}_A$ everywhere. 
Furthermore, there is a pronounced Gribov copies problem. The MCG 
gauge fixing functional has in fact many local maxima, and in practice only a 
local maximum can be achieved. The result after projection 
can be strongly dependent on the chosen maximum. 

All this is in fact inexorably connected with the physics of the problem. 
General non-Abelian configurations can have $U_p\sim 1$ everywhere, but 
the bond variables gradually wandering all over the group over long 
distances. In fact, as we argued, this is precisely why smooth extended 
vortices can survive at large beta. The MCG attempts to put every bond 
as closely to an element of the center as possible, and largely 
compress a thick vortex to a thin. It is to be expected that in 
general this cannot be achieved everywhere without 
gauge fixing ambiguities and Gribov problems. 

Numerical demonstration of the Gribov copies problem was given in 
\cite{KT4}. Starting from configurations fixed in the Lorentz gauge and 
then going to the MCG produces on average a maximum higher than starting from 
random gauge and then going to MCG. The resulting picture in the two   
cases is dramatically different. In the latter case the results of 
\cite{Get} are reproduced. In the former case there is essentially 
complete loss of string tension, while there is a drop of only about 
40\% in the density of P-vortices indicating that they cannot be associated 
with thick vortices contributing to the string tension, but only with 
short thin vortices.  

It is also known \cite {Get} that slight local smoothing of the $SU(2)$ 
configurations also causes considerable decrease in the center projected 
string tension. This again indicates ambiguities 
in associating thick vortices with P-vortices that are introduced with  
expanded cores and additional  smoothness over longer distances 
in the configurations. 

In conclusion, gauge fixing can be a useful way of isolating vortices, 
but clearly further work is needed. A more sophisticated approach is 
called for which exhibits the vortex cores (topological obstructions) 
as an intrinsic property of the gauge field, and hence independent of 
the particular gauge fixing procedure adopted. A promising proposal 
along these lines was made in \cite{forc}.  (Further discussion 
can be found in \cite{lat99}.)

\section{Necessary Condition for Confinement at Weak Coupling} 

The numerical results above  indicate that only the fluctuations 
between different $\pi_1(SU(N)/Z(N))$ homotopy sectors  
are responsible for the asymptotic string tension. Fluctuations among 
the same sector become irrelevant for large enough loops. 
Numerically, a rather delicate near cancellation between the 
different sectors occurs, resulting in area-law 
for the Wilson loop expectation (as oppposed to 
an exponentially larger perimeter-law result) at weak coupling. 
This suggests that eliminating fluctuations between different sectors 
will result in vanishing string tension, i.e. loss of 
confinement at weak coupling. Since thick vortices are precisely 
the configurations allowing jumps between different sectors as 
the continuum limit is approached, this implies that their 
presence is a necessary condition for confinement.  

A relevant rigorous result was in fact already obtained long ago in Ref. 
\cite{Y}. There it was shown that in the presence of constraints 
eliminating thick vortices completely winding around the lattice 
with periodic boundary conditions, the electric-flux free energy 
order parameter \cite{tH} in $SU(N)$ LGT exhibits non-confining 
behavior at arbitrarily weak coupling. 

The electric-flux free energy gives an upper bound on the 
Wilson loop \cite{TY}. To exhibit non-confining behavior 
for the Wilson loop itself, one needs a lower bound. 
In \cite{KT3} we recently obtained the following rigorous result.   
Consider the expectation of the Wilson loop in the presence of 
constraints that eliminate from the functional measure all 
configurations that can represent thick vortices linking with 
the Wilson loop, i.e. allow the Wilson loop to fluctuate 
into the nontrivial $\pi_1$ homotopy sectors.  
 
Then for sufficiently large $\beta$, and dimension 
$d\geq 3$ the so constrained Wilson loop expectation $W[C]$, 
exhibits perimeter law, i.e. 
there exist constants $\alpha,\ \alpha_1(d),\ \alpha_2(d)$ 
such that  
\beq 
W[C] \,\geq \, \alpha \exp\left(\, - \alpha_2\,e^{-
\alpha_1\beta}\,|C|\,\right) \quad.\label{len}
\eeq 
Here $|C|$ denotes the perimeter length of the loop $C$. 
In other words, the potential between two external quark sources 
is nonconfining at weak coupling. 

In \cite{KT3} only the $SU(2)$ case is treated explicitly. 
The result is proven for a variety of actions. 
One class of actions considered is given by: 
\beq 
A_p(U) = \beta\,\tr\,U_p + \lambda\,\mbox{sign}\,(\tr\,U_p) \;,
\eeq 
where $0\leq\lambda<\infty$ extrapolates between the standard 
Wilson action ($\lambda =0$) and the `positive plaquette action' 
model ($\lambda \to\infty$). Another choice is:  
\beq 
A_p(U_p) = \beta\;\tr\,U_p + \ln(\,\theta(\,|\tr\,U_p| - k\,)\,)\;,  
\eeq  
i.e. Wilson action with an excised small `equatorial' strip 
in $SU(2)$ of width $k$ such that $k\beta$ large as $\beta$ becomes 
large; e.g. $k$ a small constant, or $k\sim 1/\beta^{1/2}$. 
All these actions have the same naive continuum limit 
and expected to be in the same universality class \cite{Mit}.  
Choice of different actions serves to emphasize that the result is 
independent of the particular choice of YM action latticization. 

It should be emphasized that the constraints do {\it not} eliminate 
thin vortices. This is achieved by employing the $SO(3)\times Z(2)$ 
formulation \cite{MP}, \cite{T} of the $SU(2)$ LGT. All constraints depend 
only on the $SO(3)$ coset variables, and thus any $Z(2)$ plaquette 
fluxes on thin vortices remain unaffected. We refer to Ref. \cite{KT3} which 
contains detailed explicit derivations.

\section{Sufficiency Condition - Lower Bound on the 
String Tension by the Excitation Probability for a Vortex} 

As already mention an upper bound on the Wilson loop is given by the 
electric-flux free energy order parameter \cite{TY}. This quantity is the 
$Z(N)$ Fourier transform of the magnetic-flux free energy \cite{tH}.  
The magnetic flux free energy order parameter is defined    
as the ratio  $Z(z)/Z$ of the partition function 
with a `twisted' action to that with the 
original (untwisted) action. The `twist'   
inserts a nontrivial element $z\in Z(N)$, i.e. a discontinuous gauge 
transformation, in the action on every plaquette of a $(d-2)$-dim 
topologically nontrivial coclosed set of plaquettes $S^\ast$ 
(closed 2-dim surface of dual plaquettes on the dual lattice in $d=4$) 
with periodic boundary conditions. Thus $\ln (Z(z)/Z)$  
gives the free energy cost for exciting a vortex 
completely winding in $(d-2)$ spacetime directions 
around the lattice, and is also referred to as the 
vortex free energy. 
(Alternatively, 
one may consider appropriate fixed boundary conditions in the remaining 
two spacetime directions so that the winding vortex again remains 
trapped \cite{MP}.)   
The upper bound on the Wilson loop in terms of the $Z(N)$ Fourier 
transform of such `vortex containers' \cite{TY}, \cite{MP}, \cite{Yo} 
implies area-law  only if the 
vortex free energy remains finite in the large volume limit (in the 
Van Hove sense). Now to cancel a cost proportional to $L^{(d-2)}$ ($L$   
lattice linear length), the system must respond by 
spreading the discontinuous gauge transformation on $S^\ast$ in 
the two transverse directions, i.e. the vortex free energy will 
remain finite only if the expectation for exciting 
an {\it arbitrarily} long,  {\it thick} vortex 
remains finite. This provides then a sufficiency condition for 
confinement.

Recently, we have obtained an alternative lower bound on 
the string tension for $SU(2)$ which can be expressed directly in terms of 
the 't Hooft loop expectation (magnetic-disorder parameter) \cite{tH}. 
The 't Hooft operator amounts to a source exciting a  
$Z(2)$ monopole current on a coclosed set of cubes (closed loop of dual 
bonds on the dual lattice), and forming the coboundary of 
a set of plaquettes $S^\ast$ (forming the boundary of a $(d-2)$-dim 
surface of dual plaquettes) representing the attached Dirac sheet. 
The operator inserts a twist $(-1) \in Z(2)$ on each plaquette 
in $S^\ast$. In our case the monopole `loop' is 
taken to be the minimal coclosed 
set of cubes consisting of the $2(d-2)$ cubes 
sharing a given plaquette $p$. The set $S^\ast$ attached to it
winds around the lattice in the $(d-2)$ perpendicular directions.      
Again, to cancel a cost proportional to $L^{(d-2)}$, 
the system must respond by spreading a discontinuous 
gauge transformation on $S^\ast$ in the two transverse directions, i.e. 
the operator gives the expectation for exciting a thick vortex 
`punctured' by a short monopole loop. 
Now the presence of the `puncture' by the small monopole loop 
(site of the 't Hooft loop source) is a purely 
local effect that can be extracted with 
fixed action cost (at finite lattice spacing). 
Shrinking the monopole loop to a point gives then  the magnetic flux 
free-energy observable $Z(z)/Z$.

It is known that, for $SU(N)$,  individual configurations exist giving 
vanishing  vortex energy cost. The non-Abelian nature of the group 
is crucial for their existence.  No construction of a finite measure 
contribution, hence no proof at the 
nonperturbative level is available though.  

\begin{figure}[tb]
\begin{minipage}{7cm}
{\ }\hfill\epsfxsize=6cm
\epsfysize=5.5cm\epsfbox{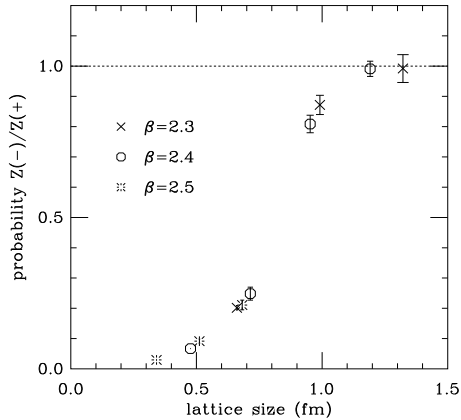}\hfill{\ }\\
\caption{Vortex probability (magnetic-flux 
free-energy) vs. lattice size} 
\label{vfe}
\end{minipage}
\end{figure}
\vspace{-0.3cm}

\section{Measurement of the Vortex Free Energy} 
\label{Vex} 
In the absence of an analytical proof, we have resorted to 
numerical evaluation of the magnetic flux free energy 
(vortex free energy) for $N=2$.    

Measurement was performed by combining Monte Carlo simulation with 
the multihistogram method of Ref. \cite{FS}. The method  was used 
in \cite{HRR} to compute the free energy of a $Z(2)$ monopole 
pair as a function of the pair's separation. 
The method consists roughly of looking at the probability distribution 
of the energy along the twist (all other variables integrated out). This probability is reconstructed by combining histograms of the energy along 
the twist obtained from several simulations at different values of the 
coupling along the twist. The method tends to be computationally 
expensive. The result of our computation is shown in figure \ref{vfe}. 
The lattice spacings are $a=0.119$ fm and $a=0.085$ fm for $\beta=2.4$ 
and $\beta=2.5$, respectively. As expected by physical reasoning, not 
only does the vortex free energy cost remain finite as 
the lattice volume grows, but it tends to 
zero, i.e. the weighted probability for the presence of a vortex 
goes to unity for sufficiently large lattice. This reflects 
the exponential spreading of color-magnetic flux in a confining phase.

\section{Conclusions} 
\label{concls}

Simulations show that the full string tension is accounted for 
by center fluctuations of the Wilson loop insensitive to short 
distance details. The result is robust under local smoothings of 
configurations, consistent with the picture of thick vortex 
configurations in the vacuum being responsible for 
confinement at weak coupling. 

The method of gauge fixing for associating  vortices  in the full theory 
with vortices in center-projected $Z(N)$ configurations 
(P-vortices) is, upon closer inspection, a rather tricky proposition. 
The common implementation (maximal center gauge) suffers from 
pronounced Gribov and smoothing problems. 
Clearly a more sophisticated approach is needed that 
manifestly does not depend on the particular gauge fixing procedure 
adopted. Work along these lines is being currently pursued.

Elimination of thick vortex configurations in the vacuum allowing 
the Wilson loop to fluctuate into different $\pi_1(SU(N)/Z(N))$ 
homotopy sectors has recently been rigorously shown to lead to loss 
of confinement at arbitrarily weak coupling. It is an old result 
that this also holds true for the electric-flux free energy order 
parameter. In other words, the presence of vortices is 
a necessary condition for confinement.  

The numerical simulations indicate that in fact the presence of 
vortices is both a necessary and 
sufficient condition. Again analytical arguments relate 
the existence of nonzero string tension directly to 
the nonvanishing of the excitation probability of a 
sufficiently thick vortex in the vacuum. A numerical 
evaluation of the vortex free energy presented here shows that 
this probability is indeed equal to unity. This indicates that 
the vacuum indeed exhibits a `condensate' of thick vortices.

\section{Acknowledgments} 

We are grateful to the organizers for their invitation and 
warm hospitality in Dubna, and for conducting such a successful,  
interactive workshop. We thank the participants for many  
discussions. The work of T.G.T. was supported by FOM, and 
of E.T.T. by NSF-PHY 9819686.

\end{document}